\documentclass{article}
\usepackage{amsmath}
\usepackage{mathtools}
\usepackage{graphicx}
\usepackage{subfigure}
\usepackage{authblk}
\usepackage{hyperref}
\usepackage{etoolbox}
\usepackage{lmodern}
\usepackage{tikz}
\usepackage[T1]{fontenc}
\usepackage{CJK}
\usepackage{algorithm}
\usepackage{algorithmic}
\usepackage{longtable}
\usepackage{braket}
\usepackage[utf8]{inputenc}
\usepackage[font=small,labelfont=bf]{caption} 
\usepackage{geometry}
\title{Qubit coupled cluster singles and doubles variational quantum eigensolver ansatz for electronic structure calculations}
\author[1]{Rongxin Xia}
\author[1]{Sabre Kais \thanks{kais@purdue.edu}}
\affil[1]{Department of Chemistry, Department of Physics and Astronomy, and Purdue Quantum Science and Engineering Institute,
Purdue University, West Lafayette, Unites States}
\date{}

\input{Qcircuit}
\begin{document}
\setcounter{MaxMatrixCols}{20}
\setlength{\parindent}{0pt}
\setlength{\parskip}{0.5\baselineskip}
\maketitle
\vspace{-8ex}

\begin{abstract}
Variational quantum eigensolver (VQE) for electronic structure calculations is believed to be one major potential application of near term quantum computing. Among all proposed VQE algorithms, the unitary coupled cluster singles and doubles excitations (UCCSD) VQE ansatz has achieved high accuracy and received a lot of research interest. However, the UCCSD VQE based on fermionic excitations needs extra terms for the parity when using Jordan-Wigner transformation.  Here we introduce a new VQE ansatz based on the particle preserving exchange gate to achieve qubit excitations. The proposed VQE ansatz has gate complexity up-bounded to $O(n^4)$ where $n$ is the number of qubits of the Hamiltonian. Numerical results of simple molecular systems such as  BeH$_2$, H$_2$O, N$_2$, H$_4$ and H$_6$ using the proposed VQE ansatz gives very accurate results within errors 
about $10^{-3}$ Hartree. 
\end{abstract}

\maketitle


\section{Introduction}

Quantum computing has been developing rapidly in recent years as a promising new paradigm for solving many problems in science and engineering. One major potential application of quantum computing is  solving quantum chemistry problems \cite{cao2019quantum} such as electronic structure of molecules, which has received a lot of research interest and achieved a big success in both algorithmic development and experimental implementation. The early development of electronic structure calculations was based on the quantum phase estimation algorithm developed by Kitaev  \cite{kitaev1995quantum}, Abrams and Lloyd \cite{abrams1997simulation} and used to find spectrum of simple molecular systems  \cite{aspuru2005simulated,wang2008quantum,daskin2011decomposition,daskin2012universal,kais2014introduction,bian2019quantum}. More recently, hybrid classical-quantum algorithms have been developed such as the  variational quantum eigenslover (VQE)  \cite{peruzzo2014variational, yung2014transistor,barends2015digital, mcclean2016theory} and quantum machine learning techniques  \cite{xia2018quantum} for electronic structure calculations.  Moreover, many experiments have been conducted on quantum computers to show that electronic structure calculations of simple molecules are possible on current Noisy Intermediate-Scale Quantum (NISQ) devices  \cite{preskill2018quantum,o2016scalable,kandala2017hardware}. 

One of the most promising quantum algorithms to perform electronic structure calculations is based on unitary coupled cluster \cite{bartlett2007coupled} singles and doubles (UCCSD), which implements the quantum computer version of UCCSD as the VQE ansatz  \cite{peruzzo2014variational,romero2018strategies,barkoutsos2018quantum} to calculate the ground state from a Hartree-Fock reference state. The results from UCCSD VQE achieve high accuracy  \cite{cao2019quantum, romero2018strategies, shen2017quantum, grimsley2019trotterized}. However,  the gate complexity for first order trotterization UCCSD VQE is up-bounded to $O(n^5)$ \cite{ cao2019quantum, romero2018strategies} using Jordan-Wigner transformation where $n$ is the number of qubits of the Hamiltonian. This makes it difficult to implement on current NISQ devices. Some strategies developed may be used to reduced the complexity, for example, the ordering and  parallelization techniques in \cite{poulin2014trotter} can reduce the circuit depth by $O(n)$ \cite{romero2018strategies} and low-rank factorization  \cite{motta2018low} can reduce the gate complexity to $O(n^4)$.  Here we introduce a new VQE ansatz based on the particle preserving exchange gate \cite{barkoutsos2018quantum,mckay2016universal} to achieve qubit excitations, which has gate complexity up-bounded to $O(n^4)$  and has comparable accuracy compared to first order trotterization UCCSD VQE. By reducing the gate complexity, QCCSD VQE ansatz -- qubit coupled cluster singles and doubles (QCCSD) VQE ansatz, might be more favorable for current NISQ devices. 

The rest of the paper is organized as follows: The first section gives a brief introduction to the method of UCCSD VQE ansatz. Then we give a detailed description of QCCSD VQE ansatz. We also show QCCSD VQE is a simplified version of the first order trotterization UCCSD VQE. Finally, we give the numerical simulation results of BeH$_2$, H$_2$O, N$_2$, H$_4$ and H$_6$ using first order trotterization UCCSD VQE and QCCSD VQE ansatz. 

\section{UCCSD VQE}
The electronic structure Hamiltonian can be written in second quantization as:

\begin{equation}
   H = h_0 + \sum_{ij}h_{ij} a_i^{\dagger}a_j +  \frac{1}{2}\sum_{ijkl}h_{ijkl} a_i^{\dagger}a_j^{\dagger}a_ka_l
\end{equation}

where $h_0$ is the nuclear repulsion energy,  the one-electron integrals $h_{ij}$ and the two-electron integrals $h_{ijkl}$ can be calculated by orbital integrals. Using Jordan-Wigner transformation we can rewrite the Hamiltonian in the Pauli matrices form:
\begin{equation}
     H = \sum_{i,\alpha} a_{\alpha}^i \sigma_{\alpha}^i + \sum_{ij, \alpha\beta} b_{\alpha\beta}^{ij} \sigma_{\alpha}^i\sigma_{\beta}^j + ...
\end{equation}

where ${a_{\alpha}^i, b_{\alpha\beta}^{ij}}$ are general coefficients and  $\sigma$ are  Pauli matrices $\sigma_x$, $\sigma_y$, $\sigma_z$ and  2$\times$2 identity matrix. 

In unitary coupled clustered single double excitations, we can calculate the ground state from the Hartree-Fock reference state by excitation operators of the form:

\begin{equation}
    |\phi\rangle = e^{T(\vec{\theta}) - T^{\dagger}(\vec{\theta})}|\phi_{HF}\rangle
\end{equation}

where $T(\vec{\theta}) = T_1(\vec{\theta}_1)  + T_2(\vec{\theta}_2)$ is the excitation operator, $|\phi_{HF}\rangle$ is the Hartree-Fock reference state and $\vec{\theta}$ is the set of adjustable parameters. The single excitation operator can be written as $T_1(\vec{\theta}_1) = \sum_{i,j}\theta_{ij}a_i^{\dagger}a_j$ and the double excitation operator can be written as $T_2(\vec{\theta}_2) = \sum_{i,j,k,l}\theta_{ijkl}a_i^{\dagger}a_j^{\dagger}a_ka_l$. We can minimize $\langle \phi|H|\phi\rangle$ to get the ground state energy by optimizing $\vec{\theta}$.

Considering an $n$ qubits Hamiltonian, the number of spin orbitals is $n$ and the total number of excitation terms in $T$ is $O({N_{occ} \choose 2}\times{N_{virt} \choose 2})$, where $N_{occ}$ is the number of occupied orbitals, $N_{virt}$ is the number of virtual spin orbitals. $n = N_{occ}+N_{virt}$ is the number of qubits of the Hamiltonian or the total number of spin orbitals. 

The first order trotterization UCCSD operator can be written as:

\begin{equation}
    e^{T(\vec{\theta}) - T^{\dagger}(\vec{\theta})} \approx \prod_{i,j}e^{\theta_{ij}(a_i^{\dagger}a_j-a_j^{\dagger}a_i)}\times\prod_{i,j,k,l}e^{\theta_{ijkl}(a_i^{\dagger}a_j^{\dagger}a_ka_l -  a_l^{\dagger}a_k^{\dagger}a_ja_i)}
\end{equation}

To map the first order trotterization UCCSD to quantum computer, we use same transformation, Jordan-Wigner transformation, as we do for the Hamiltonian to transform creation and annihilation operators into Pauli matrices. Each term in equation (4) can be implemented as unitary quantum gates by Jordan-Wigner transformation. Since the cost of Jordan-Wigner transformation for each term is $O(n)$ \cite{cao2019quantum}, the gate complexity for the first order trotterization UCCSD VQE is $O({N_{occ} \choose 2}\times{N_{virt} \choose 2}\times n) < O(n^5)$ using Jordan-Wigner transformation \cite{cao2019quantum,romero2018strategies}. 

UCCSD VQE has shown high accuracy in electronic structure calculations \cite{cao2019quantum, romero2018strategies, shen2017quantum, grimsley2019trotterized}. However, one problem of the UCCSD VQE is the large complexity.  The first order trotterization UCCSD VQE has up-bounded $O(n^4)$ terms and $O(n^5)$ gate complexity using Jordan-Wigner transformation. Here, we propose a new coupled cluster singles and doubles VQE ansatz using the particle preserving exchange gate \cite{barkoutsos2018quantum,mckay2016universal}. The gate complexity of QCCSD ansatz scales as $O({N_{occ} \choose 2}\times{N_{virt} \choose 2}) < O(n^4)$. In the numerical simulations, we show that QCCSD ansatz can achieve comparable accuracy to the first order trotterization UCCSD VQE. 

\section{QCCSD VQE ansatz}
After Jordan-Winger transformation, each qubit represents whether the corresponding spin orbital is occupied or not. When qubit $i$ is in $|0\rangle$, spin orbital $i$ is not occupied and when qubit $i$ is in $|1\rangle$ spin orbital $i$ is occupied. Thus we can write down a particle preserving exchange gate $U_{ex}$ \cite{barkoutsos2018quantum,mckay2016universal} between two qubits as:

\[
U_{ex}(\theta)=
\begin{bmatrix}
1 & 0 & 0 & 0\\
0 & cos\theta & -sin\theta &0\\
0 & sin\theta & cos\theta & 0\\
0 & 0 & 0 & 1\\
\end{bmatrix}
\]

The particle preserving exchange gate $U_{ex}$ will not change the total number occupation when applied to arbitrary states. Suppose we have two qubits in $|10\rangle$, which represents that the first spin orbital is occupied and the second spin orbital is not occupied. If we apply $U_{ex}$ to this state we have:

\begin{equation}
    U_{ex} (\theta)|10\rangle = cos\theta |10\rangle - sin\theta |01\rangle
\end{equation}

which corresponds to a single excitation between one spin occupied and one virtual spin orbitals. 

We can also write down a particle preserving exchange gate $U'_{ex}$ between four qubits as in Fig. 1. Suppose we have four qubits in $|1010\rangle$, which represents the first and the third spin orbitals are occupied while the second and the fourth spin orbitals are not occupied. If we apply $U'_{ex}$ to this state we have:

\begin{equation}
    U'_{ex} (\theta)|1010\rangle = cos\theta |1010\rangle - sin\theta|0101\rangle
\end{equation}

which corresponds to a double excitation between two occupied and two virtual spin orbitals.

 \begin{figure}[H]
\normalsize
\[
U'_{ex}(\theta)=
\begin{bmatrix}
1 & 0 & 0 & 0 & 0 & 0 & 0 & 0 & 0 & 0 & 0 & 0 & 0 & 0 & 0 & 0 \\
0 & 1 & 0 & 0 & 0 & 0 & 0 & 0 & 0 & 0 & 0 & 0 & 0 & 0 & 0 & 0 \\
0 & 0 & 1 & 0 & 0 & 0 & 0 & 0 & 0 & 0 & 0 & 0 & 0 & 0 & 0 & 0 \\
0 & 0 & 0 & 1 & 0 & 0 & 0 & 0 & 0 & 0 & 0 & 0 & 0 & 0 & 0 & 0 \\
0 & 0 & 0 & 0 & 1 & 0 & 0 & 0 & 0 & 0 & 0 & 0 & 0 & 0 & 0 & 0 \\
0 & 0 & 0 & 0 & 0 & cos\theta & 0 & 0 & 0 & 0 & -sin\theta & 0 & 0 & 0 & 0 & 0 \\
0 & 0 & 0 & 0 & 0 & 0 & 1 & 0 & 0 & 0 & 0 & 0 & 0 & 0 & 0 & 0 \\
0 & 0 & 0 & 0 & 0 & 0 & 0 & 1 & 0 & 0 & 0 & 0 & 0 & 0 & 0 & 0 \\
0 & 0 & 0 & 0 & 0 & 0 & 0 & 0 & 1 & 0 & 0 & 0 & 0 & 0 & 0 & 0 \\
0 & 0 & 0 & 0 & 0 & 0 & 0 & 0 & 0 & 1 & 0 & 0 & 0 & 0 & 0 & 0 \\
0 & 0 & 0 & 0 & 0 & sin\theta & 0 & 0 & 0 & 0 & cos\theta & 0 & 0 & 0 & 0 & 0 \\
0 & 0 & 0 & 0 & 0 & 0 & 0 & 0 & 0 & 0 & 0 & 1 & 0 & 0 & 0 & 0 \\
0 & 0 & 0 & 0 & 0 & 0 & 0 & 0 & 0 & 0 & 0 & 0 & 1 & 0 & 0 & 0 \\
0 & 0 & 0 & 0 & 0 & 0 & 0 & 0 & 0 & 0 & 0 & 0 & 0 & 1 & 0 & 0 \\
0 & 0 & 0 & 0 & 0 & 0 & 0 & 0 & 0 & 0 & 0 & 0 & 0 & 0 & 1 & 0 \\
0 & 0 & 0 & 0 & 0 & 0 & 0 & 0 & 0 & 0 & 0 & 0 & 0 & 0 & 0 & 1 \\
\end{bmatrix}
\]
\label{fig:uex}
\caption{Matrix of $U'_{ex}(\theta)$}
\end{figure}

We can write down an operator $U$ by $U_{ex}$ and $U'_{ex}$ to achieve single and double excitations from the Hartree-Fock reference state:

\begin{equation}
    |\Phi\rangle = U(\vec{\Theta})|\phi_{HF}\rangle = \prod_{i,j} U_{ex, i,j}(\theta_{ij})\prod_{i,j,k,l} U'_{ex,i,j,k,l}(\theta_{ijkl})|\phi_{HF}\rangle
\end{equation}

$U_{ex, i,j}$ represents $U_{ex}$ between qubits $i$ $j$ where qubit $i$ represents the occupied orbital and qubit $j$ represents the virtual orbital. $U'_{ex, i,j,k,l}$ represents $U'_{ex}$ between qubits $i$ $j$ $k$ $l$ where qubit $i$, $k$ represent occupied spin orbitals and qubit $j$, $l$ represent the virtual spin orbitals. $\vec{\Theta}$ is the set of adjustable parameters. We can minimize $\langle \Phi|H|\Phi\rangle$ to get the ground state energy by optimizing $\vec{\Theta}$ .

$U_{ex}(\theta)$ and $U'_{ex}(\theta)$ can be decomposed into elementary quantum gates with gate complexity $O(1)$ because the sizes of matrices of $U_{ex}(\theta)$ and $U'_{ex}(\theta)$ are $O(1)$.  A possible decomposition of $U_{ex}(\theta)$ is by Gray code \cite{nielsen2002quantum}:

\[
\Qcircuit @C=1.5em @R=2.1em {
& \ctrl{1} & \gate{R_y(2\theta)} & \ctrl{1} & \qw \\
& \targ &  \ctrl{-1} & \targ & \qw
}
\]

\begin{figure}[H]
\normalsize
\[
\Qcircuit @C=1.5em @R=2.1em {
& \ctrl{2} & \qw & \ctrl{1} &\qw &\gate{R_y(2\theta)} &\qw & \ctrl{1} & \qw & \ctrl{2} & \qw\\ 
& \qw & \ctrl{2} &\targ &\qw &\ctrl{-1}  &\qw &\targ &\ctrl{2} & \qw & \qw\\
& \targ & \qw &\qw &\gate{X} &\ctrl{-2} &\gate{X} &\qw & \qw & \targ & \qw\\ 
& \qw & \targ &\qw &\gate{X} &\ctrl{-3} &\gate{X} &\qw & \targ & \qw & \qw\\
}
\]
\caption{Decomposition of $U'_{ex}(\theta)$ as in \cite{yordanov2020efficient}}
\label{fig:udeco}
\end{figure}
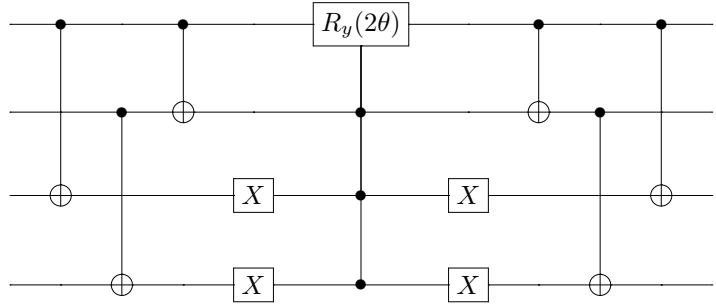

A possible decomposition of $U'_{ex}(\theta)$ as in  \cite{yordanov2020efficient} is shown in Fig. \ref{fig:udeco}. A number of groups have shown how to reduce the gate complexity of coupled cluster methods  \cite{ryabinkin2018qubit,ryabinkin2020iterative,lang2020iterative} on quantum computer and may be able to be applied to QCCSD VQE. Recently, O'Gorman et al.  \cite{o2019generalized} show that by using fermionic swap networks one can reduce the circuit depth to $O(n^{k-1})$ when implementing set of $k$ qubits gates on $n$ logical qubits, which may reduce the depth of QCCSD VQE by a factor of $n$. This is a possible future improvement but out of the scope of this paper. Recently, Yordanov et al  \cite{yordanov2020efficient} proposed a new decomposition of UCCSD VQE into two steps: first applying the qubit excitation gates, which are the particle preserving exchange gates in our QCCSD VQE used for single and double excitations though termed differently, then applying CNOT gates to include the parity information. In our simulation, $U_{ex}(\theta)$ and $U'_{ex}(\theta)$ are implemented as single unitary gates in Qiskit  \cite{aleksandrowicz2019qiskit}. 

\subsection{Excitation list selection}
One important part of the proposed VQE is to choose the excitation list, or to decide between which spin orbitals the excitation will occur. Spin preserving VQE ansatzes, which preserve the net spin magnetization $s_z$, have been widely studied  \cite{gard2020efficient,sokolov2020quantum}. We use the same strategy and choose the excitation list only allowing spin preserving exicitations. More details can be found in the Appendix A. The term complexity of our ansatz scales as $O({N_{occ} \choose 2}\times{N_{virt} \choose 2})$. The required elementary quantum gates for $U_{ex}$ and $U'_{ex}$ are both $O(1)$. Thus the gate complexity of our ansatz scales as $O({N_{occ} \choose 2}\times{N_{virt} \choose 2}) < O(n^4)$. One should note that, for linear connectivity, if no extra strategies are applied, the straightforward compilation will make the complexity of proposed QCCSD VQE scale up to $O(n^5)$. However, this complexity can be reduced by applying strategies for the compilation  as done for example in the
generalized swap network  \cite{o2019generalized}. Moreover, a recent study  \cite{gard2020efficient} shows that considering the total spin $s$ preserving may also help to achieve better accuracy, QCCSD VQE ansatz may also be able to be modified to preserve  the total spin $s$, which will be done in the future work.

\subsection{Relation to UCCSD VQE ansatz}

Here, we present that QCCSD VQE ansatz is a simplified version of UCCSD VQE ansatz. Consider a single excitation term in first order trotterization UCCSD VQE:

\begin{equation}
    e^{\theta(a_i^{\dagger} a_j-a_j^{\dagger}a_i)}
\end{equation}

Without loss of generality, we can require $i\in virt$ and $j \in occ$ where $virt$ represents virtual orbitals and $occ$ represents occupied orbitals. Furthermore we can set $i>j$ and if using Jordan-Wigner transformation we get:
 
 \begin{equation}
    e^{\frac{i\theta}{2}{\sigma_y^j\sigma_x^i\otimes_{a=j+1}^{i-1}\sigma_z^a}} e^{\frac{-i\theta}{2}{\sigma_x^j\sigma_y^i\otimes_{a=j+1}^{i-1}\sigma_z^a}}
\end{equation}

In equation (9), $\otimes_{a=j+1}^{i-1}\sigma_z^a$ counts for the parity of qubits from $j+1$ to $i-1$. If we remove the parity term we get:

\begin{equation}
    e^{\frac{i\theta}{2}{\sigma_y^j\sigma_x^i}} e^{\frac{-i\theta}{2}{\sigma_x^j\sigma_y^i}} = U_{ex,j,i}(-\theta)
\end{equation}


Thus our particle conservation exchange gate $U_{ex,j,i}$ counts for the single excitation term of qubits $j$ $i$ in first order trotterization UCCSD VQE without considering the parity of qubits from $j+1$ to $i-1$. One should be aware that if different order of spin orbitals is used, the single excitation term with parity terms removed may equal to $U_{ex}(\theta)$. Also, consider a double excitation term in first order trotterization UCCSD VQE:

\begin{equation}
    e^{\theta(a_i^{\dagger}a_j^{\dagger} a_ka_l-a_l^{\dagger}a_k^{\dagger}a_ja_i)}
\end{equation}

Without loss of generality, we can require $i<j \in virt$ and $l<k \in occ$. Furthermore we can choose the order $j>k>i>l$ and if using Jordan-Wigner transformation we get:

\begin{equation}
\begin{aligned}
     &e^{\frac{-i\theta}{8}\sigma_x^l\sigma_x^i\sigma_x^k\sigma_y^j\otimes_{a=l+1}^{i-1}\sigma_z^a\otimes_{a=k+1}^{j-1}\sigma_z^a}
     e^{\frac{-i\theta}{8}\sigma_x^l\sigma_y^i\sigma_x^k\sigma_x^j\otimes_{a=l+1}^{i-1}\sigma_z^a\otimes_{a=k+1}^{j-1}\sigma_z^a}\\
     &e^{\frac{-i\theta}{8}\sigma_x^l\sigma_y^i\sigma_y^k\sigma_y^j\otimes_{a=l+1}^{i-1}\sigma_z^a\otimes_{a=k+1}^{j-1}\sigma_z^a}
     e^{\frac{-i\theta}{8}\sigma_y^l\sigma_y^i\sigma_x^k\sigma_y^j\otimes_{a=l+1}^{i-1}\sigma_z^a\otimes_{a=k+1}^{j-1}\sigma_z^a}\\
     &e^{\frac{i\theta}{8}\sigma_x^l\sigma_x^i\sigma_y^k\sigma_x^j\otimes_{a=l+1}^{i-1}\sigma_z^a\otimes_{a=k+1}^{j-1}\sigma_z^a}
     e^{\frac{i\theta}{8}\sigma_y^l\sigma_x^i\sigma_x^k\sigma_x^j\otimes_{a=l+1}^{i-1}\sigma_z^a\otimes_{a=k+1}^{j-1}\sigma_z^a}\\
     &e^{\frac{i\theta}{8}\sigma_y^l\sigma_x^i\sigma_y^k\sigma_y^j\otimes_{a=l+1}^{i-1}\sigma_z^a\otimes_{a=k+1}^{j-1}\sigma_z^a}
     e^{\frac{i\theta}{8}\sigma_y^l\sigma_y^i\sigma_y^k\sigma_x^j\otimes_{a=l+1}^{i-1}\sigma_z^a\otimes_{a=k+1}^{j-1}\sigma_z^a}
\end{aligned}
\end{equation}

In equation (12) $\otimes_{a=l+1}^{i-1}\sigma_z^a\otimes_{a=k+1}^{j-1}\sigma_z^a$ counts for the parity of qubits from $l+1$ to $i-1$ and from $k+1$ to $j-1$. If we remove the parity term we get:

\begin{equation}
\begin{aligned}
    &e^{\frac{-i\theta}{8}\sigma_x^l\sigma_x^i\sigma_x^k\sigma_y^j}e^{\frac{-i\theta}{8}\sigma_x^l\sigma_y^i\sigma_x^k\sigma_x^j}
    e^{\frac{-i\theta}{8}\sigma_x^l\sigma_y^i\sigma_y^k\sigma_y^j}e^{\frac{-i\theta}{8}\sigma_y^l\sigma_y^i\sigma_x^k\sigma_y^j}\\
    &e^{\frac{i\theta}{8}\sigma_x^l\sigma_x^i\sigma_y^k\sigma_x^j}e^{\frac{i\theta}{8}\sigma_y^l\sigma_x^i\sigma_x^k\sigma_x^j}
    e^{\frac{i\theta}{8}\sigma_y^l\sigma_x^i\sigma_y^k\sigma_y^j}e^{\frac{i\theta}{8}\sigma_y^l\sigma_y^i\sigma_y^k\sigma_x^j}= U'_{ex, l,i,k,j}(-\theta)
\end{aligned}
\end{equation}

Thus our particle preserving exchange gate $U'_{ex,l,i,k,j}$ counts for the double excitation term of qubits $l$,$k$,$i$,$j$ in first order trotterization UCCSD VQE without considering the parity of qubits from $l+1$ to $i-1$ and from $k+1$ to $j-1$. One should be aware that if different order of spin orbitals is used, the double excitation term with parity terms removed may equal to $U'_{ex}(\theta)$. QCCSD VQE is the simplified version of first order trotterization UCCSD VQE. The reduced gate complexity of our VQE comes from removing the parity term in UCCSD VQE. Recently S. E. Smart et al  \cite{smart2020efficient} presented an efficient ansatz for two-electron system, showing that fermionic double excitations can be simplified to qubit double excitations in the two-electron system, which indicates the QCCSD VQE has the same double excitation terms as the UCCSD VQE for a two-electron system.

\section{Numerical simulation results}

In this section, we present numerical results of BeH$_2$, H$_2$O, N$_2$, H$_4$ and H$_6$ by using QCCSD VQE with gates $U_{ex}$ and $U'_{ex}$. To compare performance of QCCSD VQE, we also present the results by using first order trotterization UCCSD VQE implemented by Qiskit \cite{aleksandrowicz2019qiskit}. For each numerical simulation, the orbital integrals are calculated using STO-3G minimal basis by PySCF \cite{PYSCF} and the Hamiltonian is obtained by Jordan-Wigner transformation. The optimization is performed by the sequential least squares programming (SLSQP) algorithm \cite{kraft1988software}. The input state is the Hartree -Fock reference state and all parameters are initialized as 0 for both ansatzes. The bounds for all parameters for both ansatzes are set to $[-\pi, \pi]$. The energy thresholds for convergence is set to $10^{-6}$ Hartree with maximum 500 iterations. The noiseless simulation is done by Qiskit \cite{aleksandrowicz2019qiskit} with version 0.14.1. In the figures in this section, QCCSD VQE represent the proposed qubit coupled cluster singles and doubles VQE ansatz while UCCSD VQE represents the first order trotterization UCCSD VQE ansatz implemented by Qiskit \cite{aleksandrowicz2019qiskit}.

Complete active space (CAS) approach \cite{roos1980complete}, which divides the space to active orbitals and  inactive orbitals, has been applied to reduce the qubits of molecule Hamiltonian in quantum simulation. To investigate the effect of the size of active space, we compare the performance of QCCSD ansatz for different sizes of active spaces for the same molecule. 

\begin{figure}[H]
  \centering 
  \subfigure[The errors of ground state energies of 10 qubits BeH$_2$ Hamiltonian calculated by QCCSD VQE compared with first order trotterization UCCSD VQE.]{ 
    \includegraphics[width=2.8in]{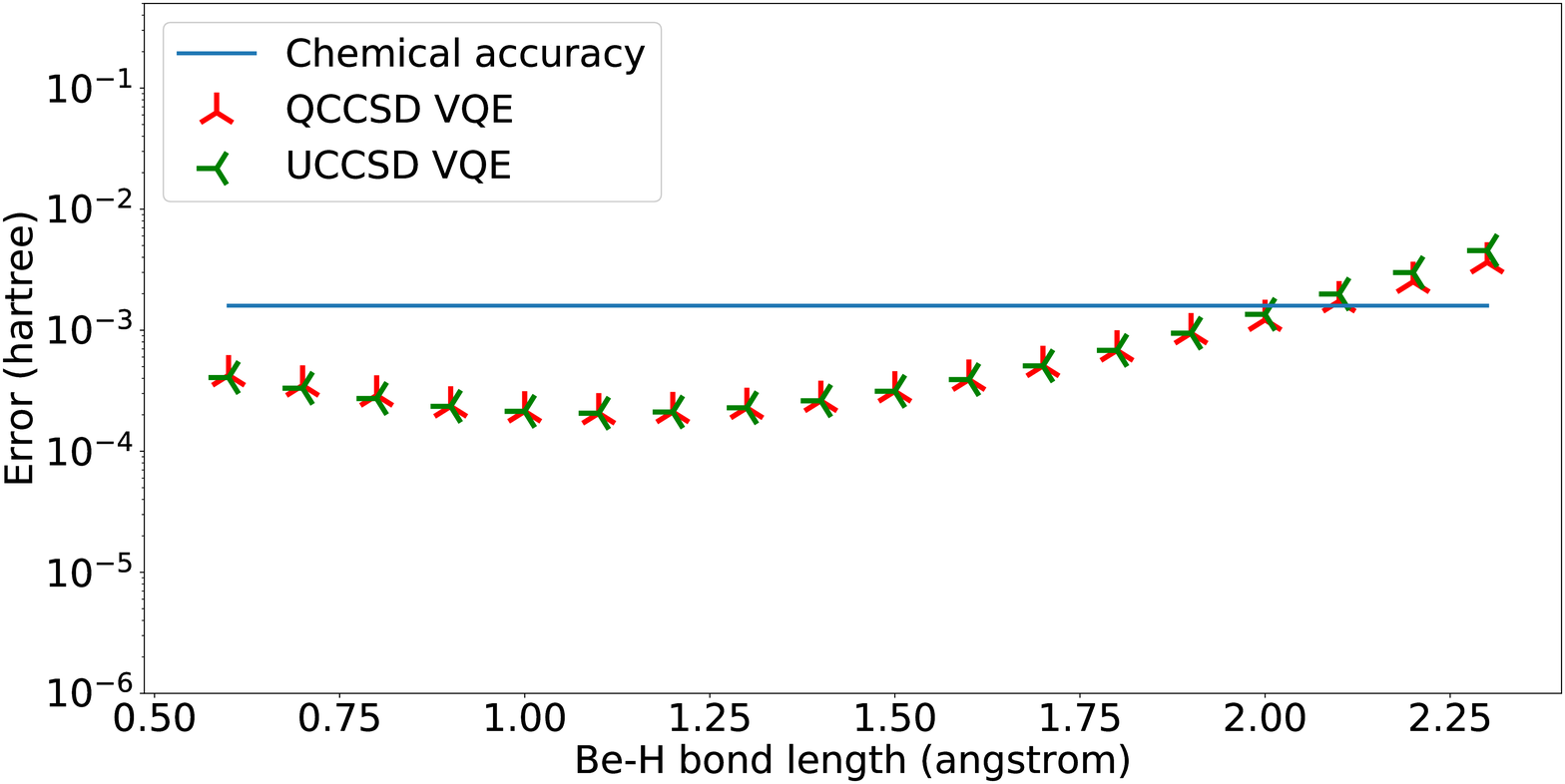}} 
  \hspace{0in} 
  \subfigure[The errors of ground state energies of 12 qubits BeH$_2$ Hamiltonian calculated by QCCSD VQE compared with first order trotterization UCCSD VQE.]{ 
    \includegraphics[width=2.8in]{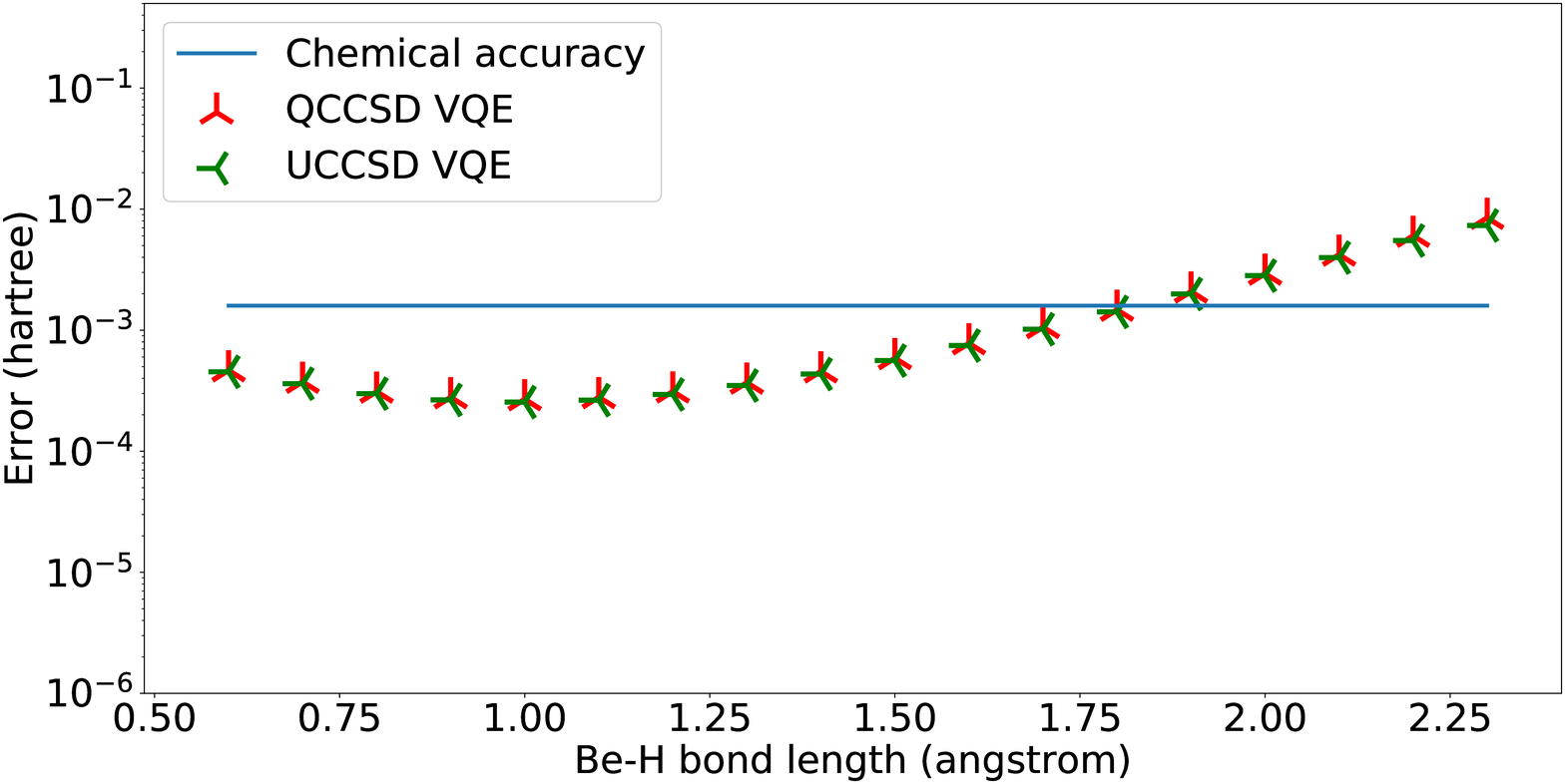}} 
  \hspace{0in} 
  \subfigure[The errors of ground state energies of 14 qubits BeH$_2$ Hamiltonian calculated by QCCSD VQE compared with first order trotterization UCCSD VQE.]{ 
    \includegraphics[width=2.8in]{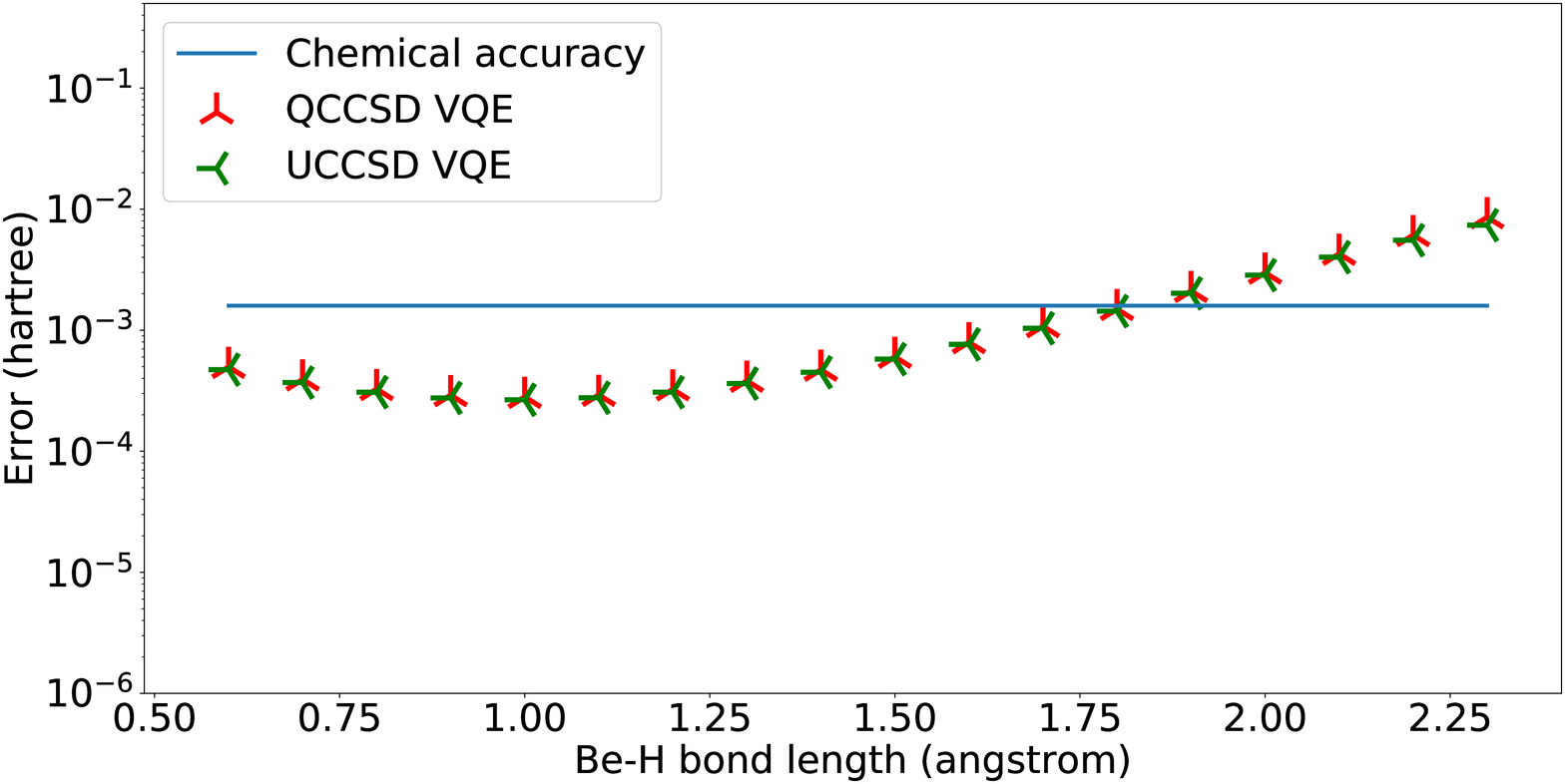}} 
  \hspace{0in} 
    \caption{VQE results of BeH$_2$ by QCCSD VQE compared with first order trotterization UCCSD VQE.}
    \label{fig:beh}
\end{figure}

For BeH$_2$ we choose three different active spaces: First 2 spin orbitals with lowest energies are always filled and first 2 spin orbitals with highest energies are always empty, corresponding to 10 active spin orbitals with 4 electrons or 10 qubits Hamiltonian. First 2 spin orbitals with lowest energies are always filled, corresponding to 12 active spin orbitals with 4 electrons or 12 qubits Hamiltonian. No spin orbitals are always filled or always empty, corresponding to 14 active spin orbitals with 6 electrons or 14 qubits Hamiltonian. We compare the errors between the ground state energies from the VQE results and the ground state energies from the diagonalization of the corresponding Hamiltonian as in Fig. \ref{fig:beh}.  We can see that, although the size of the active space is increased, our QCCSD VQE achieves similar accuracy as the first order trotterization
UCCSD VQE.

\begin{figure}[H]
  \centering 
  \subfigure[The errors of ground state energies of 10 qubits H$_2$O Hamiltonian calculated by QCCSD VQE compared with first order trotterization UCCSD VQE.]{ 
    \includegraphics[width=2.8in]{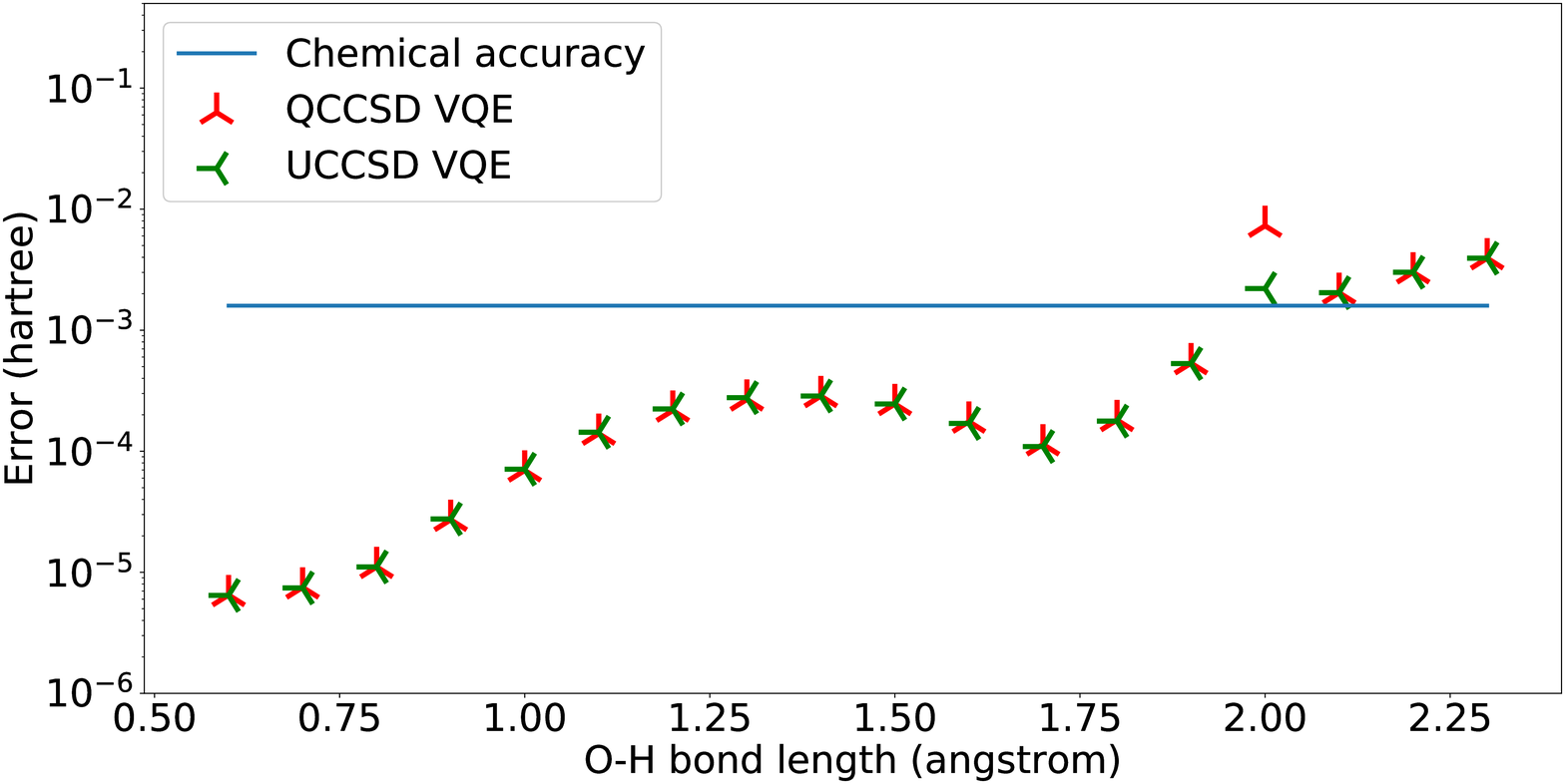}} 
  \hspace{0in} 
  \subfigure[The errors of ground state energies of 12 qubits H$_2$O Hamiltonian calculated by QCCSD VQE compared with first order trotterization UCCSD VQE.]{ 
    \includegraphics[width=2.8in]{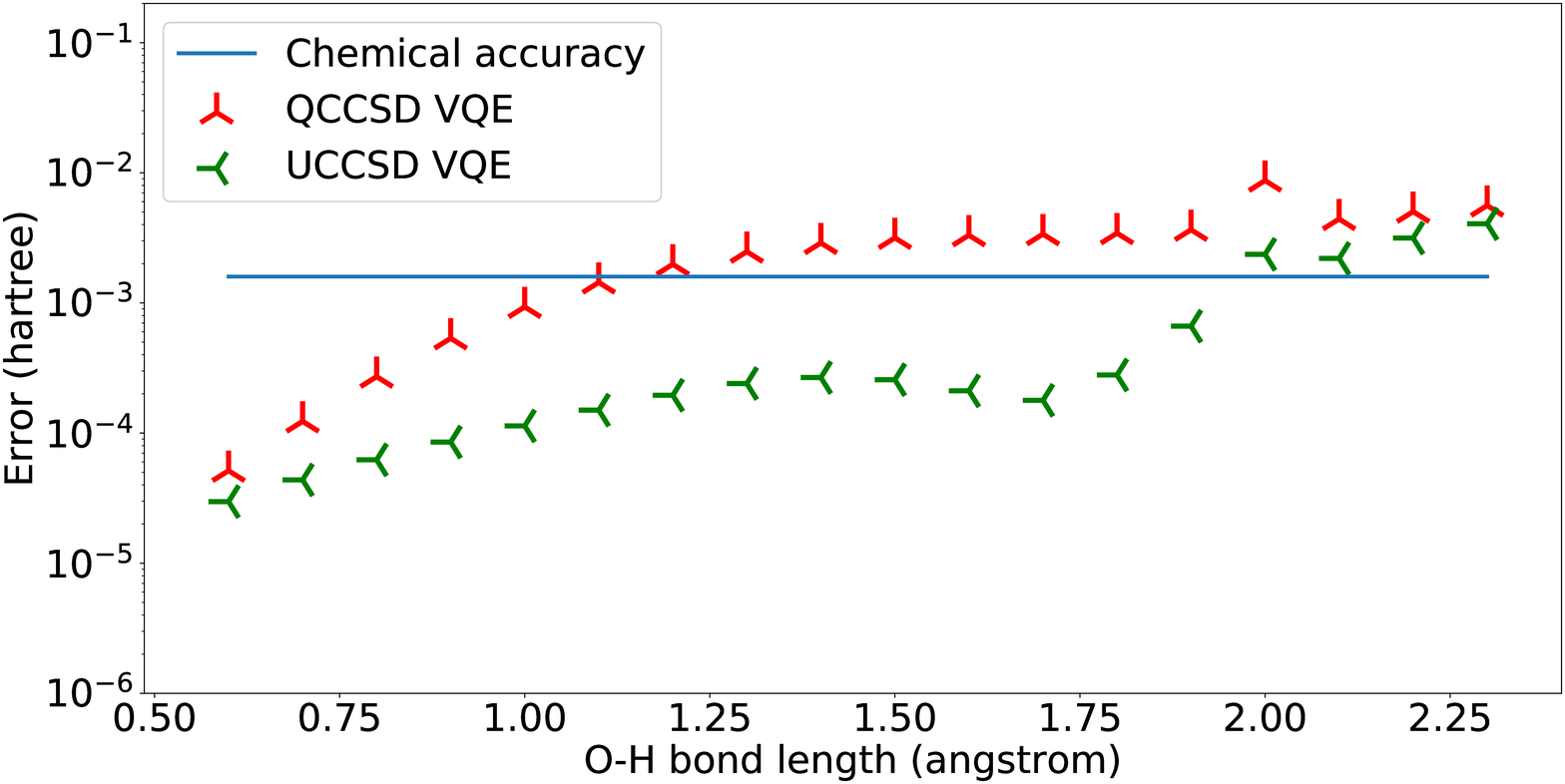}} 
  \hspace{0in} 
  \subfigure[The errors of ground state energies of 14 qubits H$_2$O Hamiltonian calculated by QCCSD VQE compared with first order trotterization UCCSD VQE.]{ 
    \includegraphics[width=2.8in]{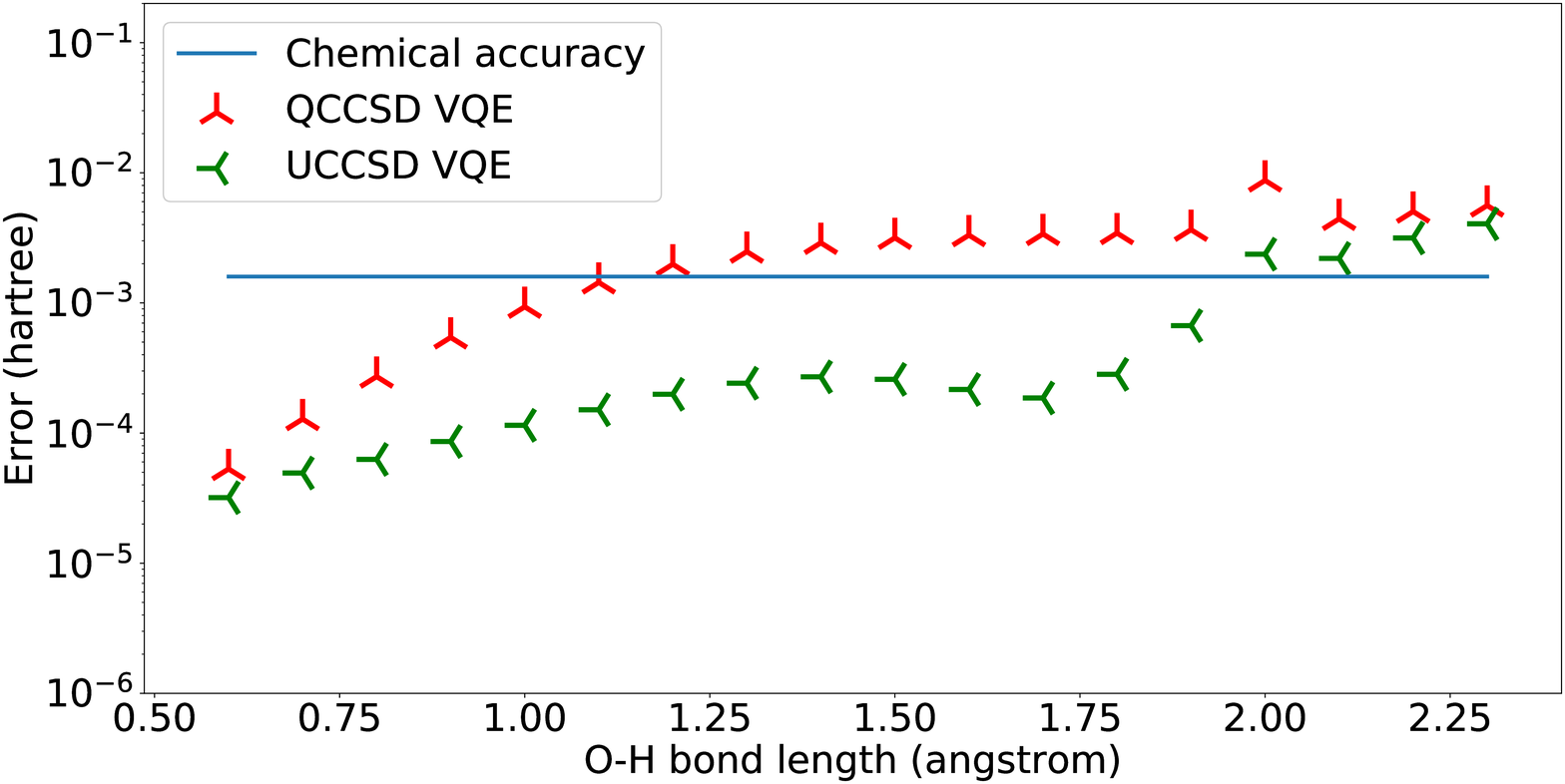}} 
  \hspace{0in} 
    \caption{VQE results of H$_2$O by QCCSD VQE compared with first order trotterization UCCSD VQE.}
    \label{fig:h2o}
\end{figure}

For H$_2$O we choose three different active spaces: First 4 spin orbitals with lowest energies are always filled, corresponding to 10 active spin orbitals with 6 electrons or 10 qubits Hamiltonian;  First 2 spin orbitals with lowest energies are always filled, corresponding to 12 active spin orbitals with 8 electrons or 12 qubits Hamiltonian and  No spin orbitals are always filled or always empty, corresponding to 14 active spin orbitals with 10 electrons or 14 qubits Hamiltonian. We compare the errors between the ground state energies from the VQE results and the ground state energies from the diagonalization of the corresponding Hamiltonian as in Fig. \ref{fig:h2o} for different Hamiltonian. For 10 qubits H$_2$ Hamiltonian our qubit coupled
cluster VQE achieves almost the same accuracy as the first order trotterization UCCSD VQE except at one point. For 12 and 14 qubits H$_2$ Hamiltonian, with increased size of active space, our QCCSD VQE performs a little worse compared to the first order trotterization
UCCSD VQE the error increased from $10^{-5}$ to $10^{-3}$ Hartree, but the error is still within or around the chemical accuracy.

\begin{figure}[H]
  \centering 
  \subfigure[The errors of ground state energies of 10 qubits N$_2$ Hamiltonian calculated by QCCSD VQE compared with first order trotterization UCCSD VQE.]{ 
    \includegraphics[width=2.8in]{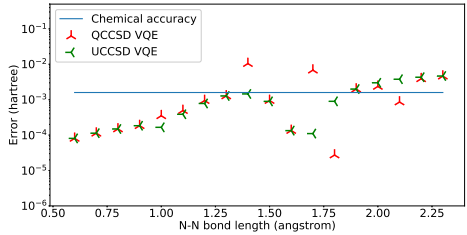}} 
  \hspace{0in} 
  \subfigure[The errors of ground state energies of 12 qubits N$_2$ Hamiltonian calculated by QCCSD VQE compared with first order trotterization UCCSD VQE.]{ 
    \includegraphics[width=2.8in]{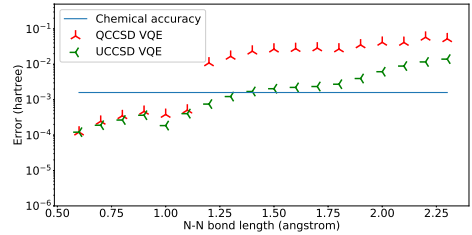}}
  \hspace{0in} 
  \subfigure[The errors of ground state energies of 14 qubits N$_2$ Hamiltonian calculated by QCCSD VQE compared with first order trotterization UCCSD VQE.]{ 
    \includegraphics[width=2.8in]{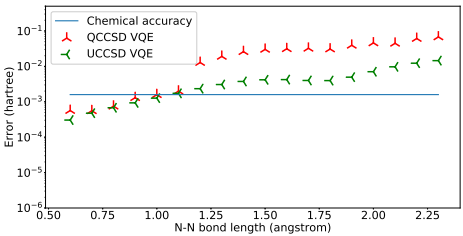}} 
  \hspace{0in} 
  \subfigure[The errors of ground state energies of 16 qubits N$_2$ Hamiltonian calculated by QCCSD VQE compared with first order trotterization UCCSD VQE.]{ 
    \includegraphics[width=2.8in]{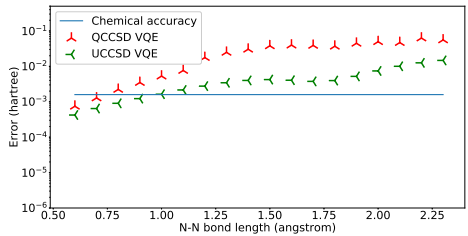}} 
  \hspace{0in} 
    \caption{VQE results of N$_2$ by QCCSD VQE compared with first order trotterization UCCSD VQE.}
    \label{fig:n2}
\end{figure}

For N$_2$ we choose four different active spaces: First 8 spin orbitals with lowest energies are always filled and first 2 spin orbitals with highest energies are always empty, corresponding to 10 active spin orbitals with 6 electrons or 10 qubits Hamiltonian; First 8 spin orbitals with lowest energies are always filled, corresponding to 12 active spin orbitals with 6 electrons or 12 qubits Hamiltonian; First 6 spin orbitals with lowest energies are always filled, corresponding to 14 active spin orbitals with 8 electrons or 14 qubits Hamiltonian. First 4 spin orbitals with lowest energies are always filled, corresponding to 16 active spin orbitals with 10 electrons or 16 qubits Hamiltonian. We compare the errors between the ground state energies from the VQE results and the ground state energies from the diagonalization of the corresponding Hamiltonian as in Fig. \ref{fig:n2} for different Hamiltonian. For 10 qubits N$_2$ Hamiltonian our QCCSD VQE achieves almost same or even better accuracy as the first order trotterization UCCSD VQE except one point. For 12, 14 and 16 qubits N$_2$ Hamiltonian, with increased size of active space, our QCCSD VQE performs worse compared to the first order trotterization UCCSD VQE. This indicates that the removal of parity terms in excitation operators may affect accuracy of the the couple cluster method for larger system size.

\begin{figure}[H]
    \centering
    \includegraphics[width=2.8in]{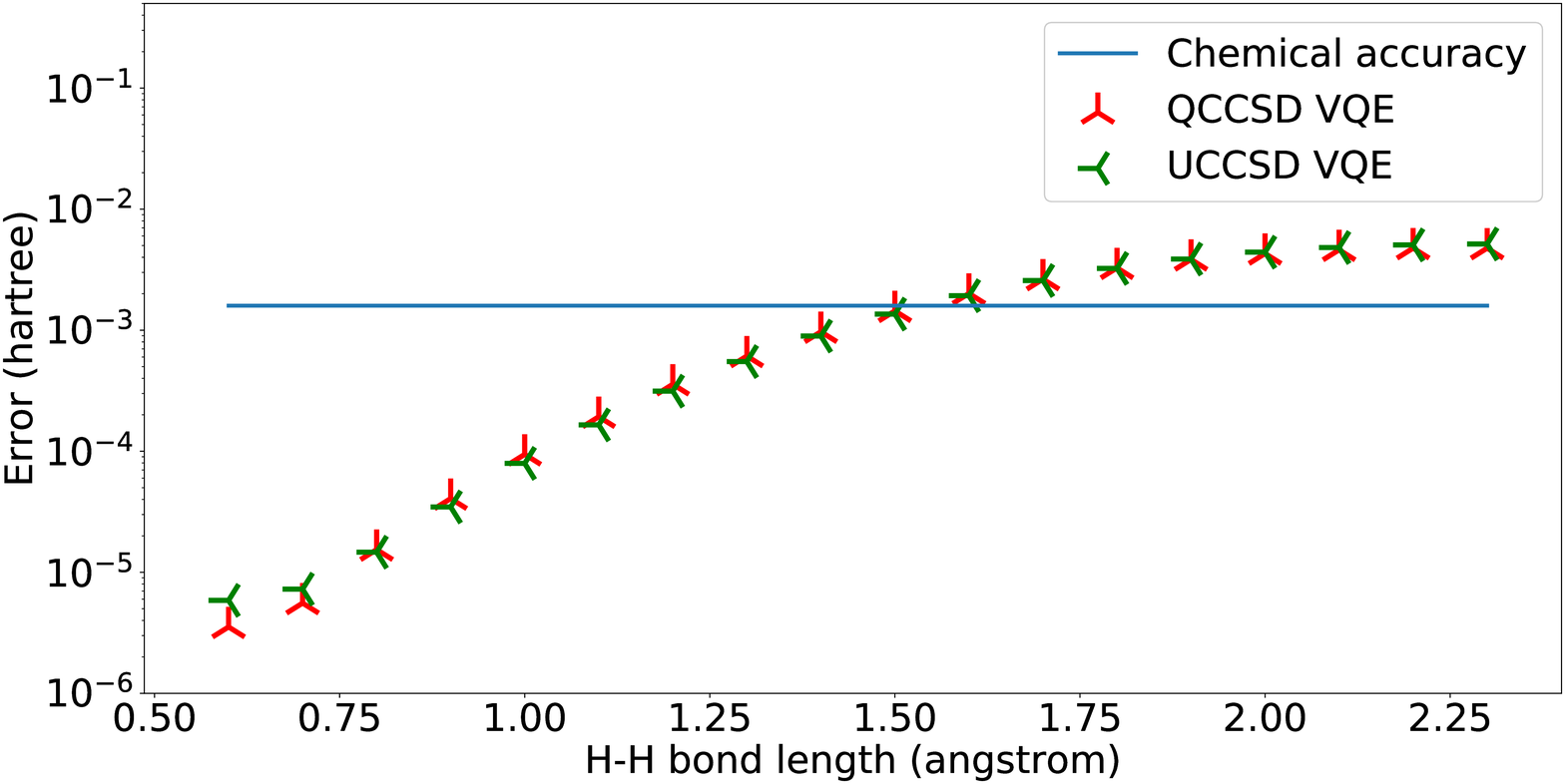}
    \caption{The errors of ground state energies of H$_4$ calculated by QCCSD VQE compared with  UCCSD VQE.}
    \label{fig:h4}
\end{figure}

For H$_4$ chain we do not have any restrictions on the spin orbitals, corresponding to 8 active spin orbitals and 4 active electrons. We also show the error between the ground state energies from VQE results and the ground state energies from the diagonalization of the corresponding Hamiltonian as in Fig. \ref{fig:h4}. our QCCSD VQE achieves the same level accuracy compared to the first order trotterization UCCSD VQE.

\begin{figure}[H]
    \centering
    \includegraphics[width=2.8in]{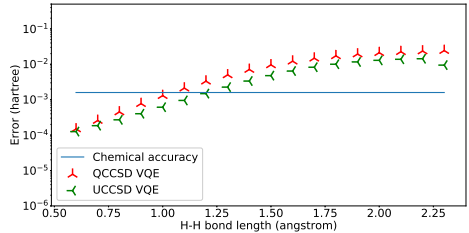}
    \caption{The errors of ground state energies of H$_6$ calculated by QCCSD VQE compared with  UCCSD VQE.}
    \label{fig:h6}
\end{figure}

For H$_6$ chain we do not have any restrictions on the spin orbitals, corresponding to 12 active spin orbitals and 6 active electrons. We also show the error between the ground state energies from VQE results and the ground state energies from the diagonalization of the corresponding Hamiltonian as in Fig. \ref{fig:h6}.  our QCCSD VQE achieves same level accuracy compared to the first order trotterization UCCSD VQE.

\section{Discussion and conclusion}

\begin{figure}[H]
    \centering
    \includegraphics[width=5in]{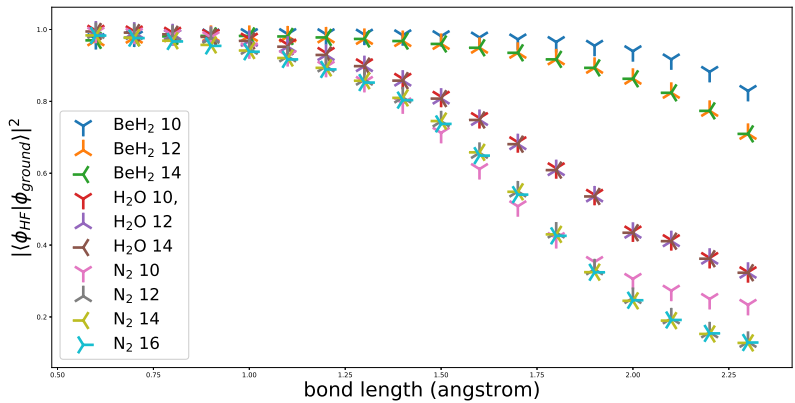}
    \caption{The overlap $|\langle\phi_{HF}|\phi_{ground}\rangle|^2$ where $|\phi_{HF}\rangle$ is the input Hartree-Fock state. and $|\phi_{ground}\rangle$ is the exact ground state obtained by diagonalization of the corresponding Hamiltonian for BeH$_2$, H$_2$O and N$_2$ of different active spaces. BeH$_2$ 10 represents the BeH$_2$ 10 qubtis Hamiltonian. BeH$_2$ 12 represents the BeH$_2$ 12 qubtis Hamiltonian. BeH$_2$ 14 represents the BeH$_2$ 14 qubtis Hamiltonian. H$_2$O 10 represents the H$_2$O 10 qubtis Hamiltonian.  H$_2$O 12 represents the H$_2$O 12 qubtis Hamiltonian.  H$_2$O 14 represents the H$_2$O 14 qubtis Hamiltonian.  N$_2$ 10 represents the N$_2$ 10 qubtis Hamiltonian.  N$_2$ 12 represents the N$_2$ 12 qubtis Hamiltonian.  N$_2$ 14 represents the N$_2$ 14 qubtis Hamiltonian.  N$_2$ 16 represents the N$_2$ 16 qubtis Hamiltonian.}
    \label{fig:overlap}
\end{figure}

In simulations, we have shown that increasing the size of active space will have little effect on the accuracy of the qubit coupled cluster VQE for BeH$_2$. our QCCSD VQE can still achieve good results for larger active space for H$_2$O but performs worse than UCCSD VQE for N$_2$. Here we present the overlap $|\langle\phi_{HF}|\phi_{ground}\rangle|^2$ where $|\phi_{HF}\rangle$ is the input Hartree-Fock state. and $|\phi_{ground}\rangle$ is the exact ground state obtained by diagonalization of the corresponding Hamiltonian for BeH$_2$, H$_2$O and N$_2$ of different active spaces. in Fig. \ref{fig:overlap}. We can see that the overlaps for BeH$_2$ with different sizes of active space are large, which may indicate very few excitation operators and small amplitudes of excitation operators are needed to approximate the exact ground state and removal of parity terms will have little effect on results. However, for N$_2$,  the overlap for N$_2$ is small when the size of active space increases and bond length is large, which may indicate that a large portion of excitation operators  and large amplitudes of excitation operators are needed to approximate the exact ground state, thus removal of parity terms may have some effects on results.

In conclusion, we have introduced a new VQE ansatz based on the particle preserving exchange gate \cite{barkoutsos2018quantum,mckay2016universal}. We have shown QCCSD VQE has reduced gate complexity from up-bounded to $O(n^5)$ of UCCSD VQE to up-bounded to $O(n^4)$ if using Jordan-Wigner transformation. In numerical simulations of BeH$_2$, H$_2$O, N$_2$, H$_4$ and H$_6$, we have shown that QCCSD VQE have achieved comparable accuracy compared to UCCSD VQE. With reduced complexity and high accuracy, QCCSD VQE ansatz might provide a new promising direction to implement electronic structure calculations on NISQ devices with chemical accuracy.

\section*{Acknowledgement}
The authors would like to thank Dr. Zixuan Hu and Teng Bian for useful discussions. This material is based upon work supported in part by the National Science Foundation under award number 1839191-ECCS and funding by the U.S. Department of Energy (Oﬃce of Basic Energy Sciences) under Award No.de-sc0019215.

\bibliographystyle{unsrt}
\bibliography{citation.bib}

\section*{Appendix}

\subsection*{Appendix.A}

As shown in \cite{grimsley2019trotterized}, the ordering of excitation operators in the Trotterized UCCSD VQE may have impact on the final results. To eliminate the effect of the operator ordering, we choose the same ordering as in the implementation of first order trotterization UCCSD VQE in Qiskit \cite{aleksandrowicz2019qiskit}. The ordering is also presented in the below algorithm.

\begin{algorithm}[H]
    \label{algorithm2}
  \caption{Qubit coupled cluster singles and doubles VQE considering spin preserving}
 \begin{algorithmic}[1]
 \FOR{orbital$_i$ in spin-up occupied orbitals}
 \FOR{orbital$_j$ in spin-up virtual orbitals}
        \STATE Construct $U_{ex}$ between qubit $i$ and $j$.
 \ENDFOR
 \ENDFOR
 
 \FOR{orbital$_k$ in spin-down occupied orbitals}
 \FOR{orbital$_l$ in spin-down virtual orbitals}
        \STATE Construct $U_{ex}$ between qubit $k$ and $l$.
 \ENDFOR
 \ENDFOR
 
 \FOR {orbital$_i$ in spin-up occupied orbitals}
 \FOR {orbital$_j$ in spin-up virtual orbitals}
\FOR {orbital$_k$ in spin-down occupied orbitals} 
\FOR{orbital$_l$ in spin-down virtual orbitals}
        \STATE Construct $U'_{ex}$ between qubit $i$ $j$ $k$ and $l$.
    \ENDFOR
\ENDFOR
\ENDFOR
\ENDFOR
    
 \FOR {orbital$_i$ in spin-up occupied orbitals}
 \FOR{orbital$_k$ in spin-up virtual orbitals}
\FOR{orbital$_j$ ($j>i$) in spin-up occupied orbitals}
\FOR{orbital$_l$ ($l>k$) in spin-up virtual orbitals}
        \STATE Construct $U'_{ex}$ between qubit $i$ $k$ $j$ and $l$.
    \ENDFOR
    \ENDFOR
    \ENDFOR
    \ENDFOR
    
 \FOR {orbital$_i$ in spin-down occupied orbitals}
 \FOR{orbital$_k$ in spin-down virtual orbitals}
\FOR{orbital$_j$ ($j>i$) in spin-down occupied orbitals}
\FOR{orbital$_l$ ($l>k$) in spin-down virtual orbitals}
        \STATE Construct $U'_{ex}$ between qubit $i$ $k$ $j$ and $l$.
    \ENDFOR
    \ENDFOR
    \ENDFOR
    \ENDFOR
 \end{algorithmic}
 \end{algorithm}
 \end{document}